\documentclass[epj]{svjour}
\usepackage{amsmath,amssymb}
\usepackage{graphicx}
\usepackage{dcolumn,mathtools,cuted}
\usepackage{color}
\usepackage{comment}
\usepackage{bbold}
\usepackage{soul,color}
\sethlcolor{white}
\usepackage{booktabs}
\usepackage{array}
\usepackage{threeparttable}
\usepackage{hyperref}
\hypersetup{
    colorlinks=true,       
    linkcolor=blue,          
    citecolor=blue,        
    filecolor=blue,      
    urlcolor=blue           
}

\begin{document}



\title{Systematic study of high-K isomers in the midshell Gd and Dy nuclei}


\author{S. K. Ghorui\inst{1}
\thanks{surja@sjtu.edu.cn}
\and C. R. Praharaj\inst{2}
\thanks{crp@iopb.res.in}
}
\institute{School of Physics and Astronomy, Shanghai Jiao Tong University, Shanghai 200240, China \and 
Institute of Physics, Bhubaneswar-751005, INDIA
}

\abstract{
High-K isomers are well known in the rare-earth region and provide unique access to the high spin structures of the nuclei. With the current interest in the study of neutron-rich rare-earth nuclei at Radioactive Ion Beam (RIB) facilities, we present here  theoretical results of the band structures of neutron-rich Gd and Dy nuclei, including the high K-isomers. Apart from the already known K-isomers, we predict some more K-isomers and these are suggested for future studies at RIB facilities. Self-consistent Deformed Hartree-Fock and Angular Momentum Projection theories are used to get the intrinsic structures, band-spectra and electromagnetic transitions probabilities of the ground band as well as bands based on isomers.
}

\maketitle


\section{Introduction}\label{intro}
Nuclear isomers \cite{Walker1999,Sun2005} are excited, metastable quantum-mechanical states of nuclei. 
The quantum degree of freedom, K, which is the angular-momentum projection onto the symmetry axis of a 
deformed nucleus, is usually a good quantum number. Because of K selection rule \cite{Bohr1975}, decay from a 
high-K state to low-K state can be hindered. Therefore, they have longer half-lives than usual excited states and form
high-K isomers. K-isomeric states appear widely throughout the nuclear chart~\cite{Walker1999,Walker2016}. 
Moreover, high-K isomers systematically occur in neutron-rich nuclei with $A\ge$150 which 
typically have deformed prolate shapes. Nuclear K-isomers decay predominantly by electromagnetic processes
($\gamma$-decay or internal conversion). There are also known instances of 
the decay being initiated by the strong interaction ($\alpha$-emission) or 
the weak interaction ($\beta$-decay or electron capture). Decay by proton or 
neutron emission, or even by nuclear fission, is possible for some isomers.
The K-isomers serve as unique access to high-spin structure of neutron-rich nuclei.
Therefore, the study of nuclear isomers, their occurrences and stability gives insight on nuclear structure 
of nuclei. The exploration of nuclei with isomeric states is also  interesting from astrophysical point
of view to understand how they affect the nucleosunthesis paths and the creation of elements heavier than Fe in the Universe \cite{Burbidge1957}.

The single-particle structure of nuclei in the deformed rare-earth region is still an interesting topic.
In the A$\sim160$ mass region nuclei, high-$\Omega$ (projection quantum number on symmetry axis) orbits are 
available near the Fermi surfaces. Therefore, nuclei in this region are prime candidates for high-K isomers~\cite{Walker1999,Walker2016,Dracoulis2016}. The decay of these high-K isomers to the ground band, which is predominantly K=0 for even-even nuclei, is hindered due to large difference in the
K-values of the isomers from the ground band.  The presence of isomeric states and hence their  decay  to low-lying states provide insight into the nuclear structure, specially collectivity in these nuclei. It has long been recognized that the nature of nuclear
collectivity at low spins depends on the number of proton
and neutron valence particles or holes outside the
closed shells~\cite{Casten1981,Casten1993}. Therefore,  the high degree of axial symmetry and large deformation are
expected for a doubly midshell nuclei. 
Thus one expects $^{170}$Dy, having large-$\Omega$ orbits near the Fermi surfaces, to be 
a good candidate for long-lived K-isomers.
An earlier calculation~\cite{Regan2002} using the total Routhian surface (TRS) and configuration constrained
potential energy surface (PES) methods predicted that pure, axially symmetric
deformed shape is  favored for mid-shell nuclei $^{170}$Dy. This is also supported by Skyrme Hartee-Fock and Projected Hartree-Fock calculations in Ref.~\cite{Rath2003}. In both the articles, a high-K isomer at low excitation energy is predicted \cite{Regan2002,Rath2003}. However, a recent measurement~\cite{Soderstrom2016} shows that the $K^\pi=6^+$ isomer 
decay hindrance factor is reduced by an order of magnitude compared to the predictions. 
S{\"o}derstr{\"o}m {\it et. al.}~\cite{Soderstrom2016} claimed that the reduced hindrance is due to 
mixing with other K-configurations.
Recent calculations using the triaxial projected shell model showed a strong correlation between the isomer hindrance and the properties of the $\gamma-$band for heavier N=104 isotones~\cite{Chen2013}. 
 In recent years many 2-qp isomers are observed experimentally for N=100-104 isotones (\cite{Simpson2009,Wang2014,Patel2014,Yokoyama2017,Ideguchi2016} and refs. therein). Spectroscopic properties of high-K isomers in deformed nuclei with $A>100$ are recently evaluated by Kondev {et. al.}~\cite{Kondev2015}. Interestingly, using isomer decay spectroscopy, Patel {et. al.} \cite{Patel2014} showed the increased quadrupole collectivity near N=100 for Sm and Gd nuclei as predicted in Refs.~\cite{Ghorui2012,Satpathy2003}.
 
 From astrophysical abundance pattern, apart from the prominent peaks at
 $A\sim130$ and $A\sim190$, there is a small bump near $A=160$. This is
 well-known as the rare-earth element (REE) peak. It has been suggested that the deformation plays important role in the formation of the REE peak~\cite{Surman1997}. Although, the structure of neutron-rich nuclei in the rare-earth region has broad range of interests, spectroscopic information are very limited. With the advancement of radioactive-ion 
 beam facilities (e.g., RIBF at RIKEN) many experiments have been performed for nuclei in this mass region \cite{Patel2014,Yokoyama2017,Soderstrom2016,Soderstrom,Gurgi2017,Wantanabe2016}. In fact, the rare-earth nuclei at N=100 and beyond are candidates for study of neutron-rich nuclei at the experimental facilities~\cite{Yokoyama}. Given the fact that neutron-rich nuclei are topic of current interest, it is proper to have theoretical results and predictions of deformations and possible K-isomers in this active region using microscopic models. These will further facilitate  
 the experimental investigations. Deformed Hartree-Fock theory \cite{Ripka1966} (based on the variational principle) for intrinsic structure  and Angular Momentum Projection \cite{Peierls1957} for restoration of rotational symmetry are very useful for such study.

Here we study the systematics of isomeric bands  and  electromagnetic properties of rare-earth nuclei employing the Deformed Hartree-Fock (DHF) and Angular Momentum (J) projection 
method \cite{Ripka1966,Praharaj1982}.
Residual interaction is included in building the deformed basis in this theory. From 
the self-consistent Hartree-Fock (HF) calculations, we build the intrinsic states of the bands by particle-hole excitations 
across the proton and neutron Fermi surfaces (the various particle-hole configurations based on HF intrinsic 
states), besides the Hatree-Fock configuration. J-projection from the deformed intrinsic configurations gives 
the spectra and electromagnetic properties of various bands \cite{Praharaj1982,Ghorui2012a}. Diagonalisation after projection can be done. This model, with the residual interaction built into the HF states, is very close to the shell model as has been shown in earlier 
studies \cite{Macfarlane1971,Khadkikar1971}.

\section{Deformed Hartree-Fock and Angular Momentum Projection Formalism}

The present model, based on quantum many-body method,  has been quite successful in explaining high spin states in the rare-earth region \cite{Ghorui2012,Rath1993,Naik2003,Ghorui2014} and also in the lighter
mass region \cite{Praharaj1988,Ghorui2011}.
In this section we briefly discuss the model used for the microscopic calculations.
More details can be found in Refs.~\cite{Ripka1966,Praharaj1982,Ghorui2012a}. 
Our model consists of self-consistent deformed Hartree-Fock mean field obtained
with a Surface Delta residual interaction and subsequent angular momentum projection
to obtain state with good angular momentum. 

The axially deformed states $|\eta m\rangle$ are expanded in the spherical basis states as follows:
\begin{equation}
|\eta m\rangle=\sum_{j}C^{\eta}_{jm}|jm\rangle
\end{equation}
where $j$ is the angular momentum of the spherical single particle state and $m$ its
projection on symmetry axis. The mixing amplitudes $C^{\eta}_{jm}$ are obtained by solving 
deformed Hartree-Fock equations in an iterative process (See Appendix~\ref{appn} for more details). When the convergence in the 
HF solutions is obtained we get deformed single particle orbits. The residual interaction
is also included self-consistently and it causes the mixing of  single-particle orbits of nucleons. The deformed HF basis is  enriched compared to the
Nilsson basis as  the $p-p$, $n-n$ and $p-n$ correlations are built in by the
inclusion of  residual
interaction in a self-consistent manner through the HF procedure (Eq.~\ref{eq:hf} in the Appendix~\ref{appn}).

 Because of mixing in the single particle orbits, the HF configurations $|\phi_K\rangle$ are
superposition of states of various $J$ values. The states of good angular momenta can
be extracted by means of projection operator \cite{Peierls1957}  

\begin{equation}
P_{K}^{JM} = \frac{2J+1}{8\pi^2}\int d\Theta \ {D_{MK}^J(\Theta)}^*
\ R(\Theta)
\label{eq_1}
\end{equation}
here $R(\Theta)$ is the rotation operator 
$e^{-i\alpha J_z}e^{-i\beta J_y}e^{-i\gamma J_z}$
and $\Theta$ represents the Euler angles $\alpha$, $\beta$ and $\gamma$. The angular momentum projection operator \ref{eq_1} restores the rotational symmetry broken during HF procedure. 
By means of angular momentum projection, various high-K configurations can be described without giving any preference to the orientation of the rotation axis. 

The Hamiltonian overlap between two states of angular 
momentum $J$ projected from intrinsic states $|\phi_{K_1}\rangle$ and  $|\phi_{K_2}\rangle$ is given by:

\begin{strip}
\begin{eqnarray}
\langle\psi_{K_1}^{J_1}|H|\psi_{K_2}^{J_2}\rangle = \frac{2J+1}{2} \frac{1}{(N_{K_1K_1}^JN_{K_2K_2}^J)^{1/2}}
{{{\int}_{0}}^{\pi}} d\beta \ sin(\beta)d_{K_1K_2}^J(\beta) 
\langle\phi_{K_1}|He^{-i\beta J_y}|\phi_{K_2}\rangle
\label{eq_2}
\end{eqnarray}
\end{strip}
\noindent with 
\begin{equation}
N_{K_1K_2}^J = \frac{2J+1}{2}{{{\int}_{0}}^{\pi}} d\beta \ sin(\beta)
d_{K_1K_2}^J(\beta)\langle\phi_{K_1}|e^{-i\beta J_y}|\phi_{K_2}\rangle
\label{eq_3}
\end{equation}

Reduced matrix elements of tensor operator $T^L$ between projected
states $|\psi_{K1}^{J_1}\rangle$ and  $|\psi_{K2}^{J_2}\rangle$ are given by
\begin{strip}
\begin{eqnarray}
\langle\psi_{K_1}^{J_1}||T^L||\psi_{K_2}^{J_2}\rangle = \frac{1}{2}
\frac{(2J_2+1)(2J_1+1)^{1/2}}{(N_{K_1K_1}^{J_1}N_{K_2K_2}^{J_2})^{1/2}}
\sum_{\mu\nu}C_{\mu\nu K_1}^{J_2LJ_1} 
{{{\int}_{0}}^{\pi}} d\beta \ sin(\beta)d_{\mu K_2}^{J_2}(\beta)
\langle\phi_{K_1}|T^L_{\nu}e^{-i\beta J_y}|\phi_{K_2}\rangle
\label{eq_4}
\end{eqnarray}
\end{strip}
\noindent where the tensor operator $T^L$ denotes electromagnetic operators of multipolarity $L$.

In general, two states $|{\psi_{K_1}^{JM}}\rangle$ and
$|{\psi_{K_2}^{JM}}\rangle$
projected from two intrinsic configurations $|\phi_{K_1}\rangle$ and
$|\phi_{K_2}\rangle$ are not orthogonal to each other even if
 the intrinsic states $|\phi_{K_1}\rangle$
and $|\phi_{K_2}\rangle$ are orthogonal. We orthonormalise them
 using following  equation
\begin{equation}
\sum_{K^{\prime}}(H^J_{KK^{\prime}}-E_JN^J_{KK^{\prime}})
b^J_{K^{\prime}}=0
\end{equation}
Here $N^J_{KK^{\prime}}$ are amplitude overlap and $b^J_{K^{\prime}}$ are
the orthonormalised amplitudes, which can be identified as
band-mixing amplitudes. The orthonormalised states are given by
\begin{equation}
\Psi^{JM}=\sum_{K}b^J_{K}\psi_{K}^{JM}
\end{equation}
With these orthonormalised states we can calculate matrix elements of various tensor operators. The mixing of intrinsic K structures can be deduced from the 
orthonomalized wave functions.

\begin{figure}[h]
\begin{center}
\includegraphics[width=0.45\textwidth]{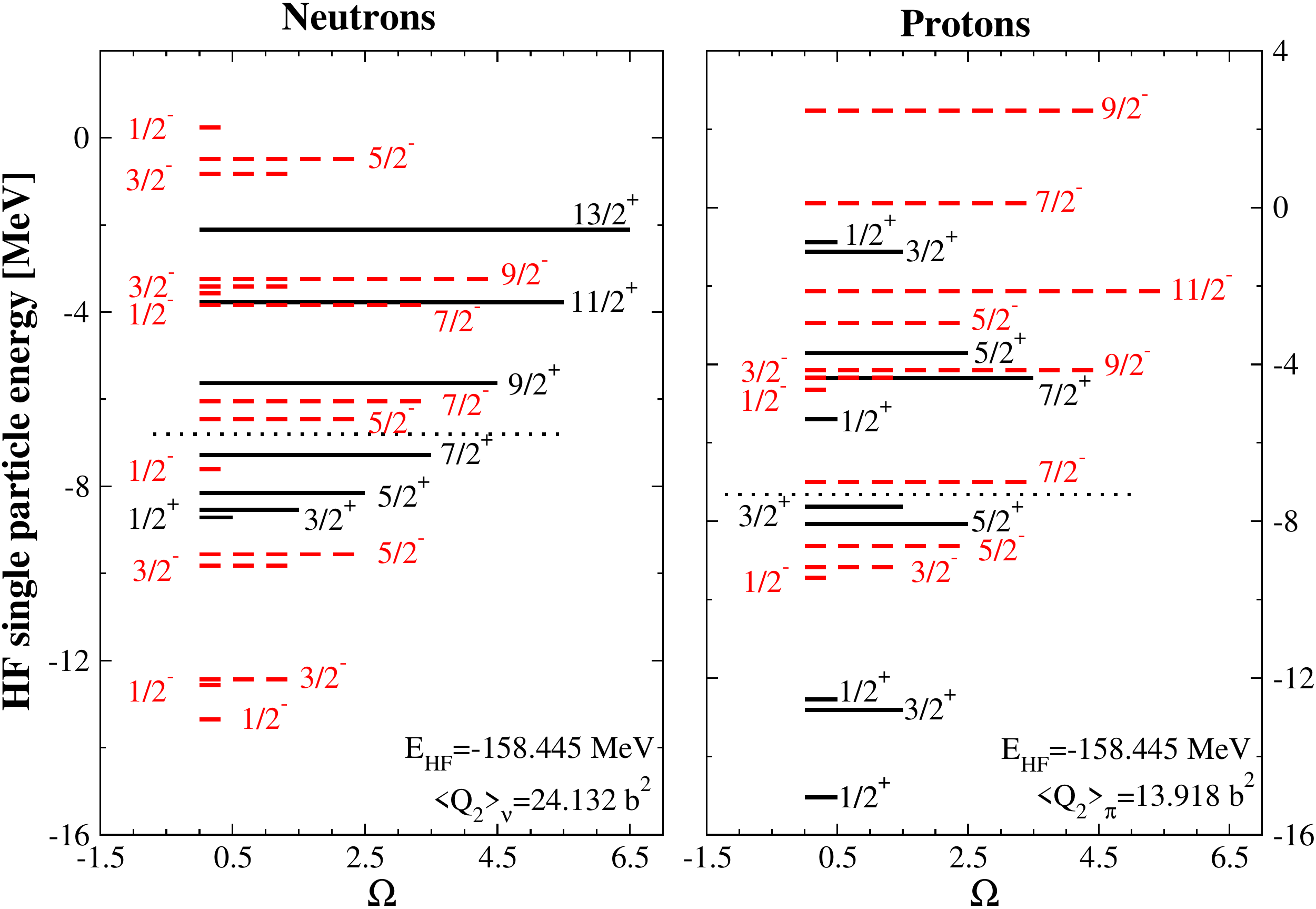}
\end{center}
\caption{\label{fig:hforb}
[color online] The prolate Hartree-Fock single-particle orbits 
 for neutrons (left panel) and protons (right panel)  are shown for  $^{168}$Dy.  
  Solid and dashed lines correspond to the positive and negative 
parity orbits, respectively. 
The length of these lines indicate the magnitude
of the z-projection ($\Omega$) of the angular momentum.
Dotted lines guide the eye to the Fermi levels. }
\end{figure}

\begin{figure}[h!]
\includegraphics[width=0.45\textwidth]{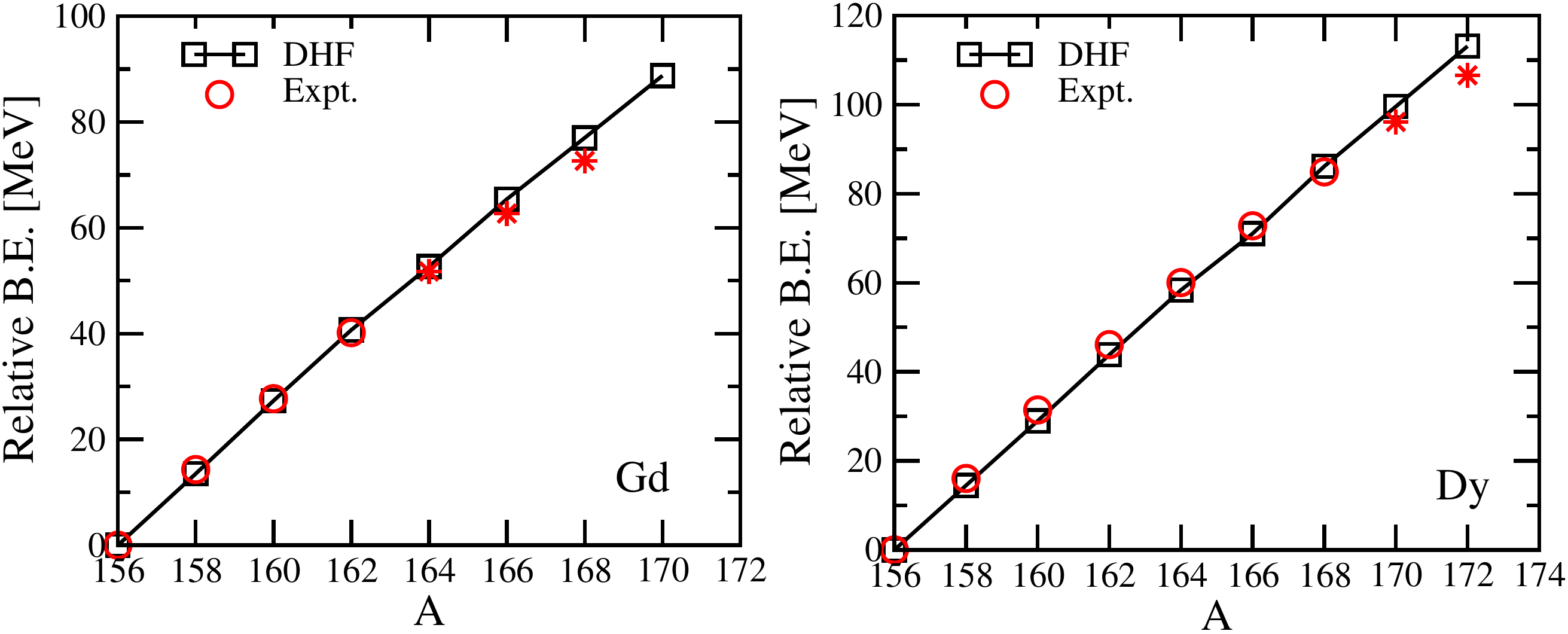}
\caption{\label{fig:be-rel}
Hartree-Fock energies of Gd (left panel) and Dy (right panel) isotopic chains calculated with DHF theory (open squares)
compared with the available experimental binding energies (circles) or extrapolated values (asterisks). The calculated and experimental
values are normalized w.r.t. A=156. Experimental and extrapolated data are
taken from \cite{ame2016}.
}
\end{figure} 

\begin{figure}[h!]
\includegraphics[width=0.45\textwidth]{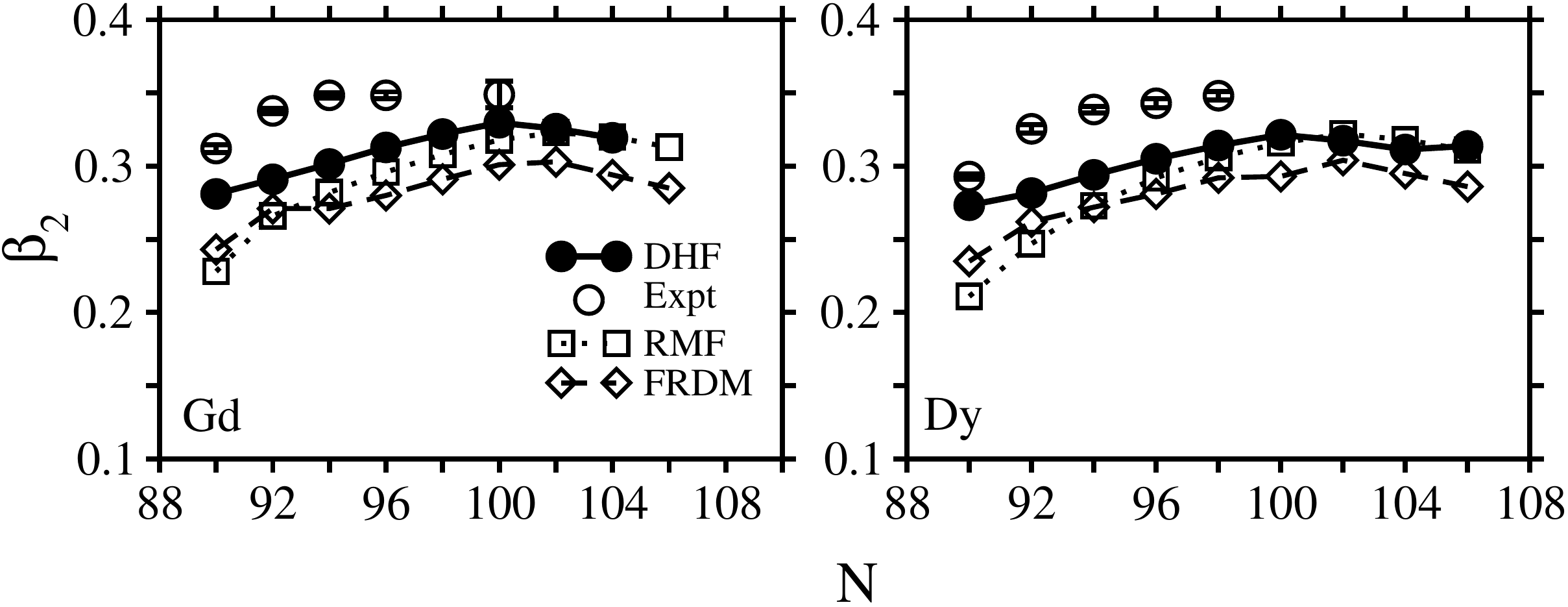}
\caption{\label{fig:beta2}
Quadrupole deformation ($\beta_{2}$) vs neutron number (N) plot for Gd and Dy nuclei. Experimental data are
taken from \cite{Raman2001,nndc}
}
\end{figure}  
\section{Results and Discussion}
In deformed (axial) Hartree-Fock and 
angular momentum projection (DHF) technique [for details see 
\cite{Praharaj1982,Ghorui2011}  and references there in]
we start with a model space 
and an effective interaction. The model space is presently
limited to one major shell  for protons and neutrons
lying outside the $^{132}$Sn core. 
 The 3$s_{1/2}$, 2$d_{3/2}$, 2$d_{5/2}$,  1$g_{7/2}$, 1$h_{9/2}$ and 1$h_{11/2}$
 proton 
states have energies  3.654, 3.288, 0.731, 0.0,   6.96 and 1.705 MeV, and the 
3$p_{1/2}$, 3$p_{3/2}$, 2$f_{5/2}$, 2$f_{7/2}$, 1$h_{9/2}$ and 1$i_{13/2}$  
neutron states have energies 4.462,   2.974,  3.432,  0.0,  0.686 and
1.487 MeV  respectively \cite{Bohr1975b,Kahana1969}.
The prolate HF calculation for the valence nucleons 
lying outside the $^{132}$Sn core is performed for both Gd and Dy isotopic chains presently studied.
 The set of  prolate deformed HF  orbits  (with well defined
$\Omega$-quantum numbers) shown in Fig.~\ref{fig:hforb}
for $^{168}$Dy forms the deformed single particle basis 
for the valence protons and neutrons. We would like to mention that
the valance spaces used here, limited to one major shell each for protons and neutrons,
are relatively small but adequate. Since the nuclei presently studied are lying near the middle of the shell so the shells above and below do not contribute much for the low-lying structures. Moreover, the important orbits which are responsible for low-lying high-K configurations are all included in the valance space employed here. So in that sense 
the present model space is adequate to describe the properties of nuclei studied here. It is worthwhile to mention that to study higher-lying structures, it would be necessary to employ
a valance space consisting of more than one major shell. We use surface delta Interaction (SDI)~\cite{Faessler1967} as the residual interaction among the active
nucleons within the valance space.The strength of the SDI  has been taken to be 0.3 MeV for $p-p$, $n-n$ and $p-n$ interactions in our calculation~\cite{Ghorui2012a,Rath1993}. The strength was fixed to
reproduce relative binding energies of nuclei in the rare-earth mass
region~\cite{Ghorui2012a}. In spite of its simple nature, this interaction gives a good description of the systematics of deformations in this mass region~\cite{Ghorui2012a,Praharaj2011}. Also
the interaction reproduces quite well the relative experimental binding energies
of the Gd and Dy isotopic chains as shown in Fig.~\ref{fig:be-rel}.

\begin{table*}[h!]
\centering
\caption{The values of HF energy (E$_{HF}$), interaction energies ($\left< V_{pp}\right>$, $\left< V_{nn}\right>$, $\left< V_{pn}\right>$), the quadrupole [Q$_{2}=\left< r^{2}Y_{20}(\theta)\right >$] and
hexadecapole [Q$_{4}=\left < r^{4}Y_{40}(\theta)\right>$] moments obtained for prolate Hartree-Fock
ground states in $^{164,166,168}$Gd and $^{166,168,170}$Dy. The multipole moments are given in units of the harmonic
oscillator length parameter, $b=0.9A^{1/3}+0.7$ fm.}
\resizebox{0.85\textwidth}{!}{
\begin{tabular}{ccccccccccccc}
\hline\hline
Z & A & K$^{\pi}$ & E$_{HF}$ &$\left< V_{pp}\right>$ & $\left< V_{nn}\right>$ & $\left< V_{pn}\right>$  &&  \multicolumn{2}{c}{Q$_2$ in $b^2$} & & \multicolumn{2}{c}{Q$_4$ in $b^4$}  \\ 
\cline{9-10}\cline{12-13}
   &     &   & [MeV] &[MeV]& [MeV]&[MeV] &&  Proton & Neutron   &&     Proton & Neutron \\
   \hline
64 & 164 & 0$^+$ &-130.962  &-20.983 & -38.154 & -112.593  &&  13.728 & 23.370 &&  12.993 & 33.338 \\
     &  166 &0$^+$ &-143.689 &-21.016 &-44.700 &-121.992  &&  13.749 & 23.997        &&  12.612 & 14.416 \\
     &  168 &0$^+$ &-155.287 &-21.031 &-49.960 &-128.970  &&  13.765 & 23.480        &&  12.379 & -6.388 \\     
66 &  166 & 0$^+$ &-144.341 &-25.418 &-38.307 &-123.017  &&  13.900 & 23.512        &&  9.098  & 30.275 \\
     &  168 &0$^+$ &-158.445 &-25.455 &-44.857 &-133.775   &&  13.918 & 24.132        &&  8.763 & 11.459 \\
     &  170 &0$^+$ &-171.859 &-25.482 &-50.142 &-142.538  &&  13.930 & 23.596        &&  8.433 & -9.122 \\     
\hline\hline
\end{tabular}
}
\label{tab:hf-prop}
\end{table*}

The nuclei studied in the present work are axially symmetric deformed nuclei and,
therefore, have good K quantum number.
Here we comment on the systematics of the proton/ neutron quasiparticle
excitations involving large $\Omega$ orbits (of different parities ) near the
respective Fermi surfaces (can be seen in Fig.~\ref{fig:hforb}). Such 
excitations are candidates for isomers, some of which have already been seen 
for nuclei in this region \cite{Ghorui2014}. The HF orbits (5/2$^-$, 7/2$^-$, 9/2$^-$) 
near the proton Fermi surface are built on $h_{11/2}$ and $h_{9/2}$ 
predominantly. Similarly, 
the HF orbits near the neutron Fermi surface are built on $i_{13/2}$ and 
$h_{11/2}$. 
The orbits near the Fermi surfaces (both proton and neutron) are
particularly important for K-isomers. 
\begin{figure*}[h!]
\begin{center}
\resizebox{0.9\textwidth}{!}{
\begin{tabular}{ccc}
\includegraphics[width=0.6\columnwidth]{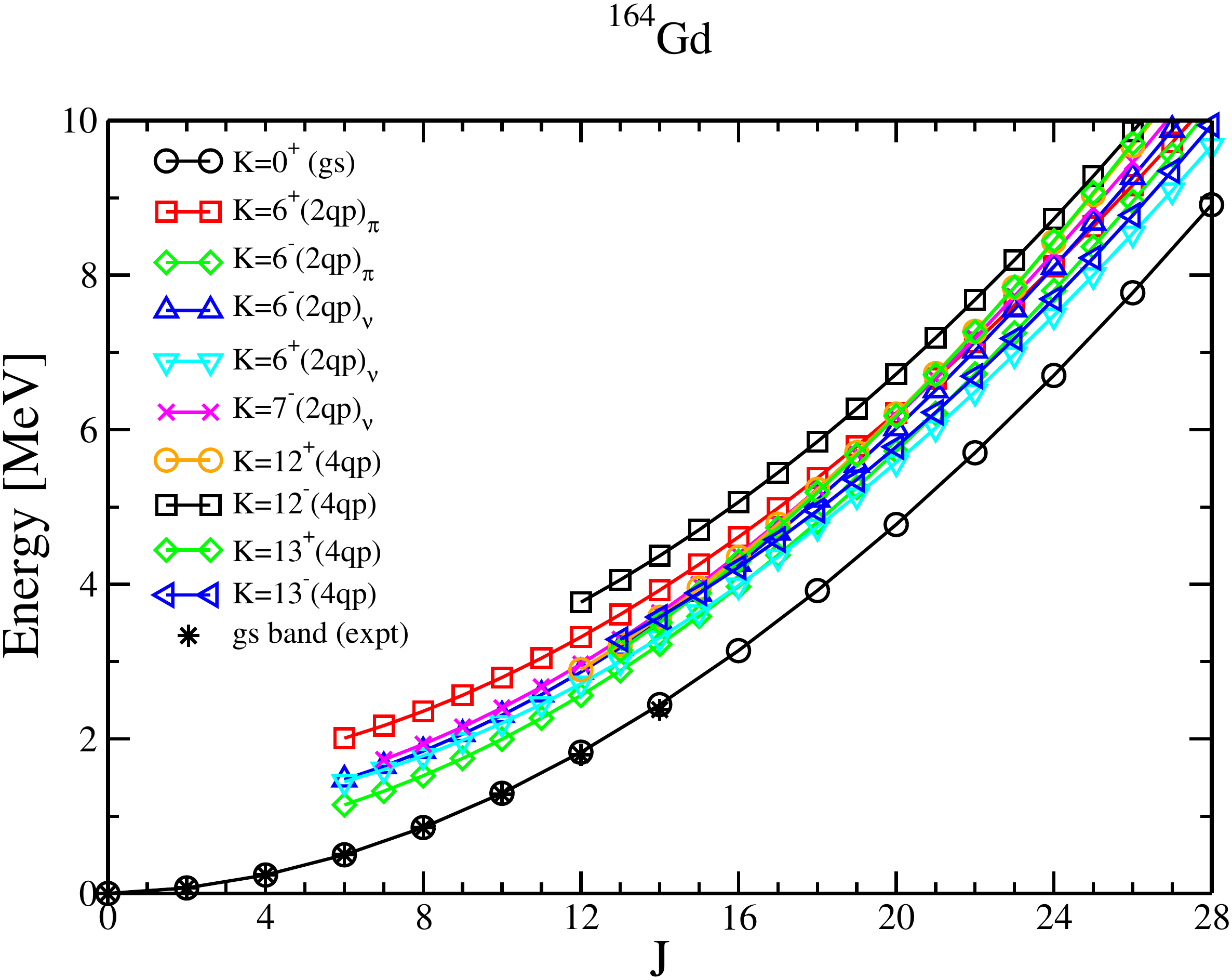} & 
\includegraphics[width=0.6\columnwidth]{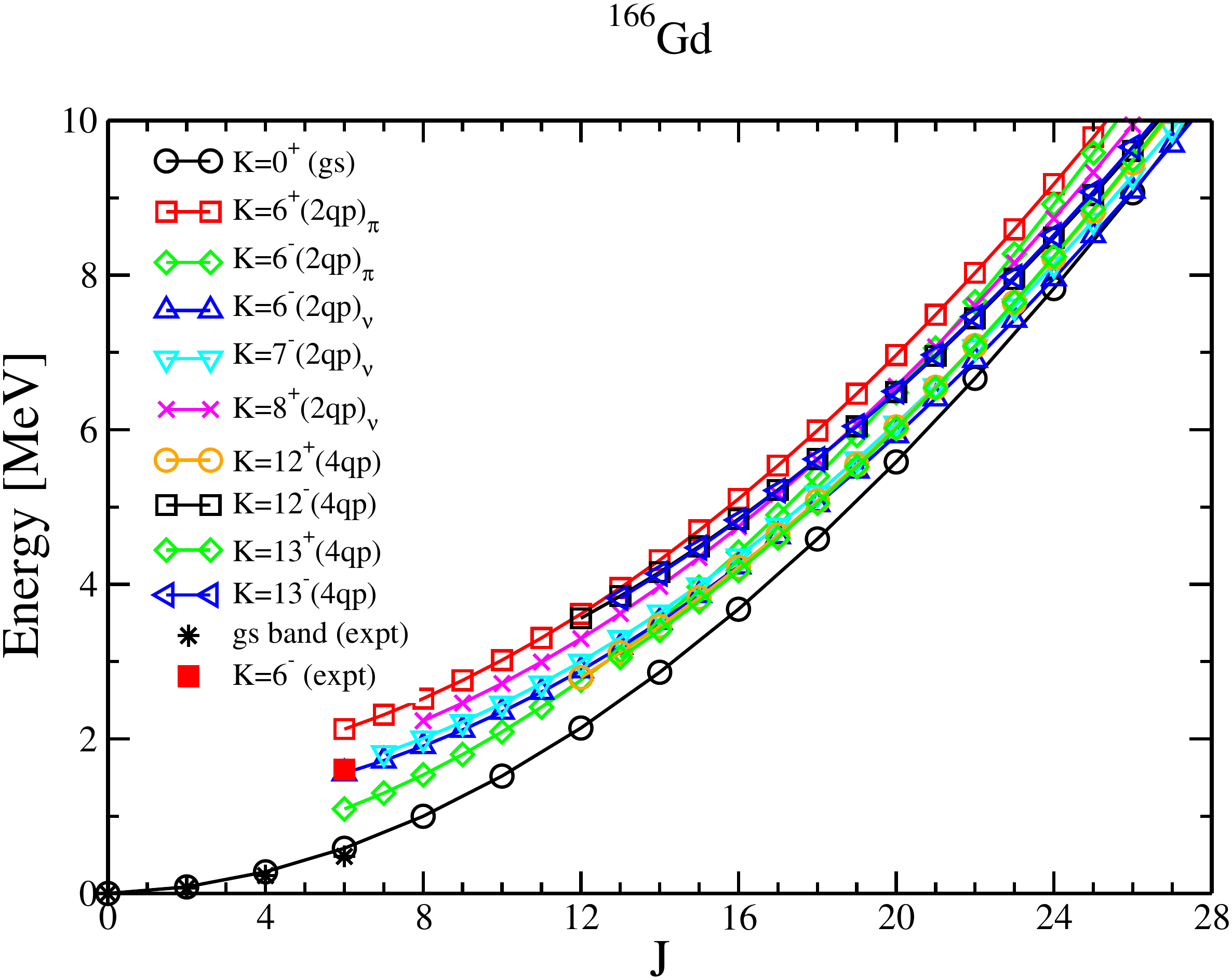} &
\includegraphics[width=0.6\columnwidth]{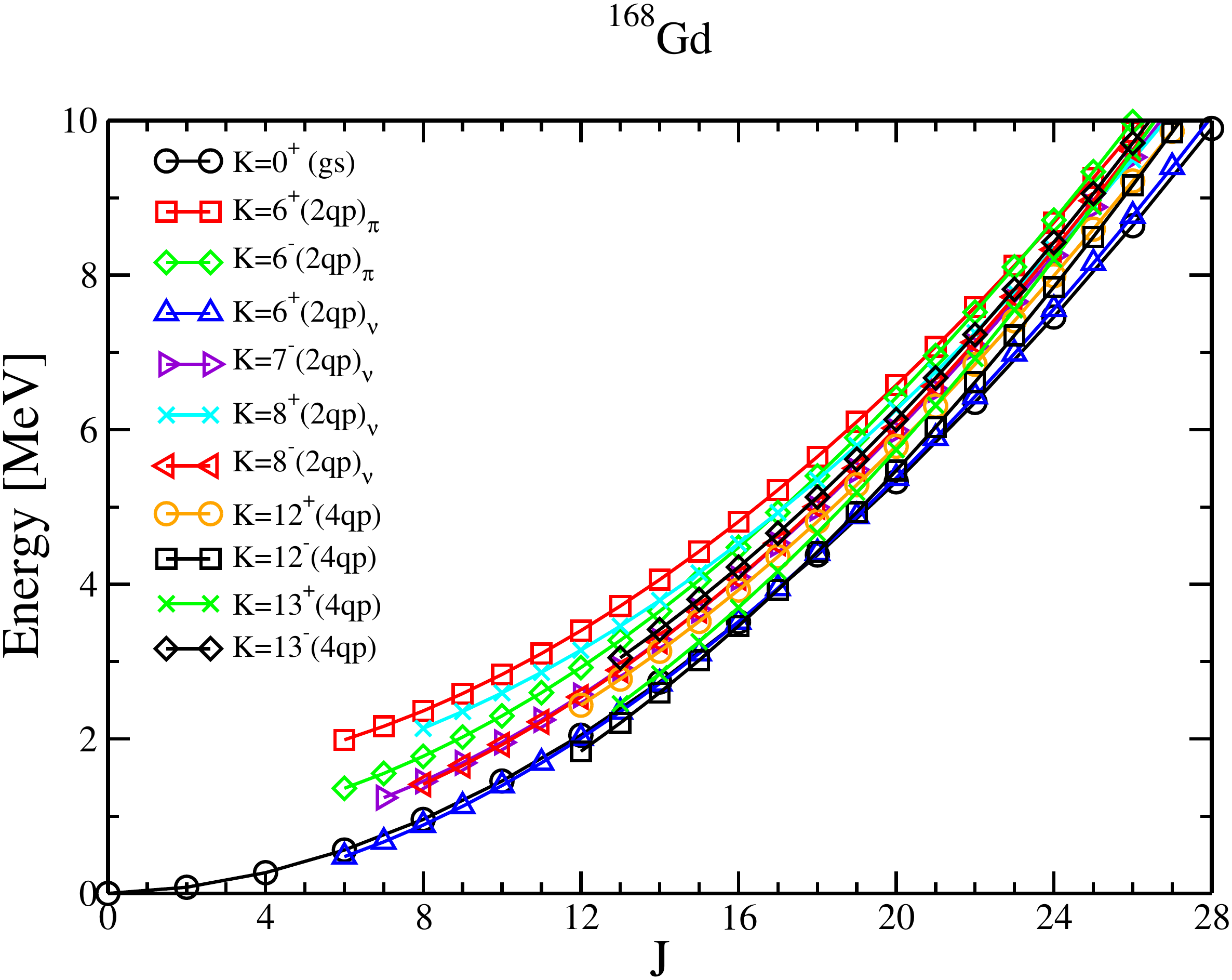} \\
(a) E vs J plot for $^{164}$Gd &
(b) E vs J plot for $^{166}$Gd &
(c) E vs J plot for $^{168}$Gd \\
& & \\
\includegraphics[width=0.6\columnwidth]{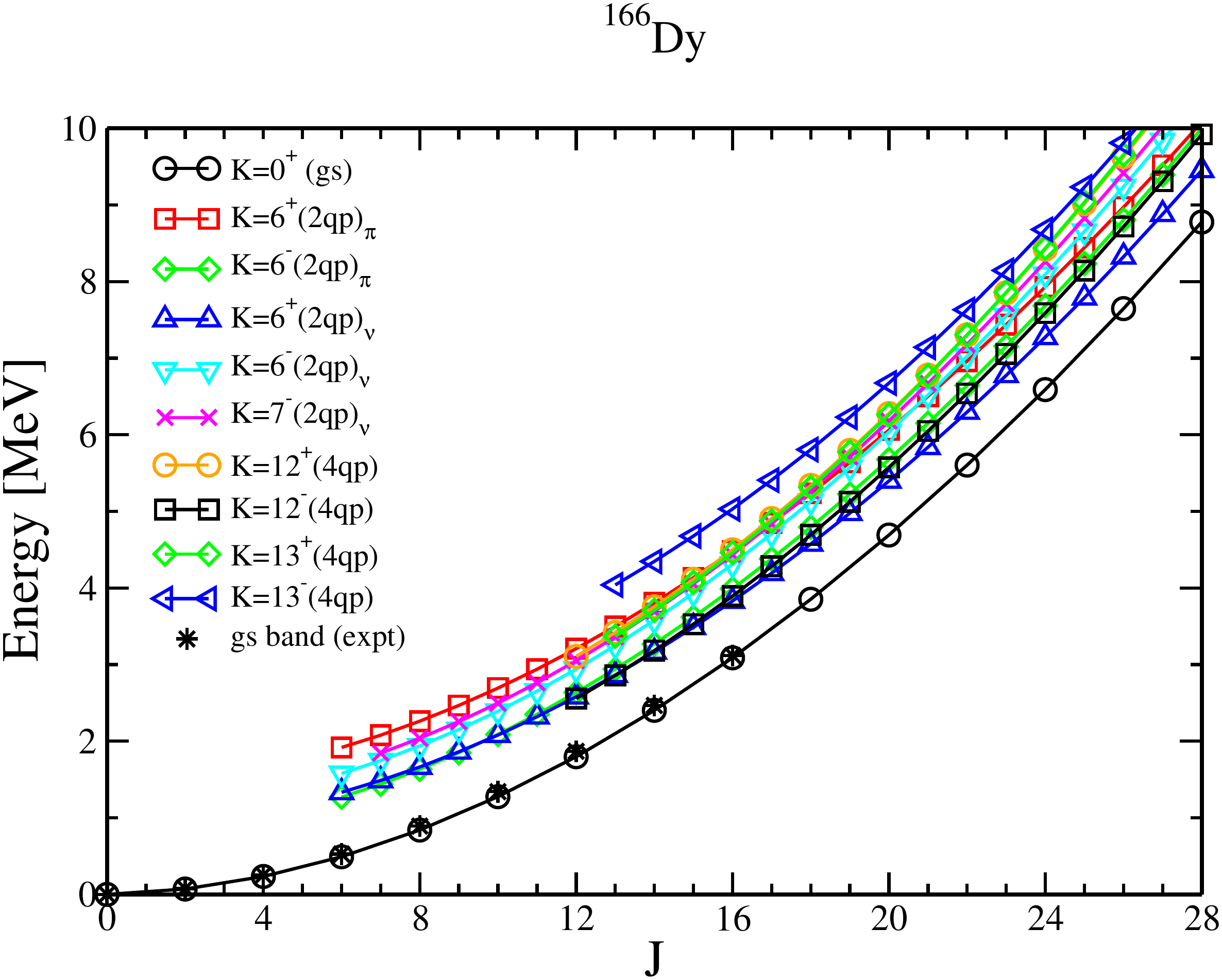} &
\includegraphics[width=0.6\columnwidth]{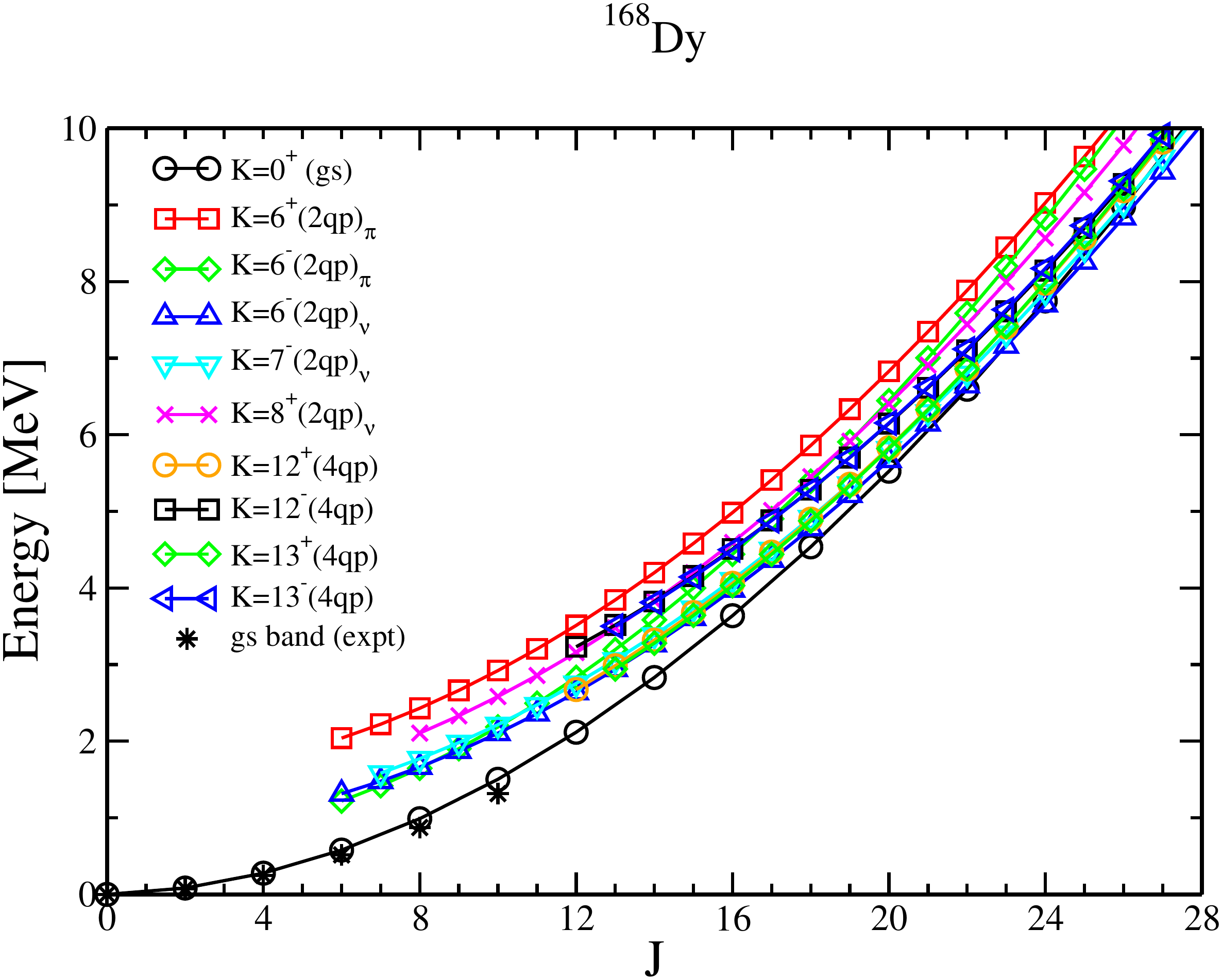} &
\includegraphics[width=0.6\columnwidth]{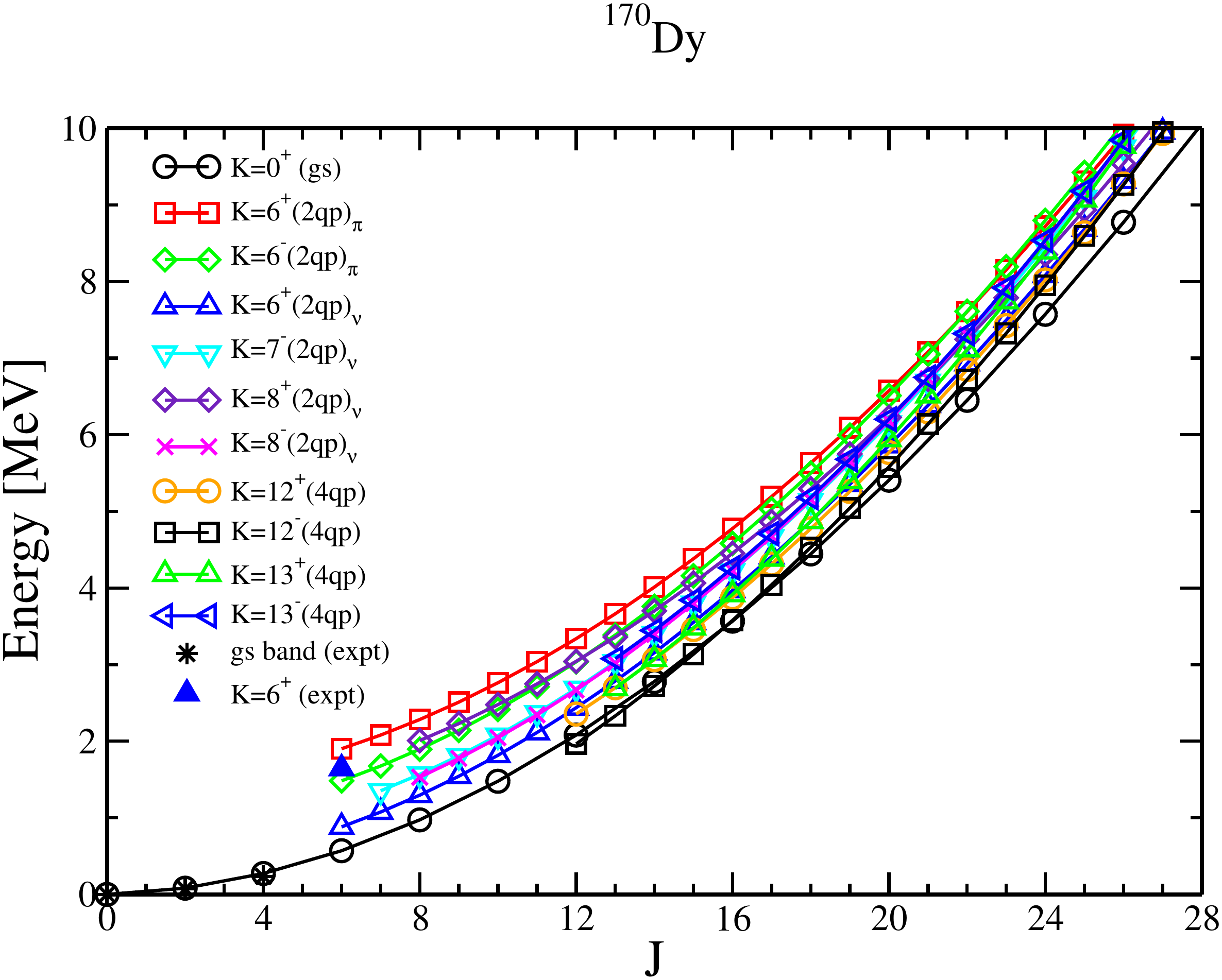} \\
(d) E vs J plot for $^{166}$Dy &
(e) E vs J plot for $^{168}$Dy &
(f) E vs J plot for $^{170}$Dy \\
\end{tabular}
}
\end{center}
\caption{\label{fig:band-diagram}Comparison of theoretical and experimental energy spectra. Experimental data are taken from 
\cite{nndc,Singh2018,Baglin2008,Baglin2010,BAGLIN2002}.}
\end{figure*}
Before going further, we want to quantify the deformation behaviour of Gd and Dy  isotopic chains.
This shows the ability of the present mean-field calculations in the description of
general behaviour of nuclei in this mass region. As proton and neutron orbits are filled up beyond the closed-shell, the
deformation increases and hence the collectivity. The maximum deformation is expected in the middle of the shell. 
The $^{170}$Dy isotope with Z=66 and N=104 has the largest number of valance particles for any isotope with A $<$ 208. 
Hence,  $^{170}$Dy is considered as the most collective nuclei in its ground state \cite{Regan2002}. However, experimental 
data are not available at present for this isotope.  In Fig.~\ref{fig:beta2}, we have plotted the quadrupole
deformation ($\beta_{2}$) parameters for N=90 to 106 isotopes of Gd and Dy. The quadrupole and hexadecapole moments 
for the prolate Hartree-Fock solutions are given in Table~\ref{tab:hf-prop}. Our DHF results for $\beta_{2}$ are 
compared with available experimental~\cite{Raman2001,nndc} as well as RMF~\cite{Lalazissis1999} and FRDM~\cite{Moller1995} calculations. 
All the theoretical values including DHF results  
 are little lower than the experimental data. But the overall trends are correctly reproduced. The $\beta_{2}$ values 
steadily increase with N and are nearly stabilized after N=96 with maximum at N$\sim$100. It shows that $\beta_{2}$ value
tends to decrease beyond N=102.  From Table~\ref{tab:hf-prop} we can see that the nuclei presently studied possess large static  ground-state quadrupole moments. 

\begin{table*}[!ht]
\centering
\caption{The dominant configuration, band-head Energy (BHE), Quadrupole moment (Q$_{S}$) and $g$-factor of the isomeric configurations.}
\begin{threeparttable}[b]
\resizebox{0.97\textwidth}{!}{
\begin{tabular}{ccccccccccc}
\hline
Isotope & K$^{\pi}$ & Configuration & & \multicolumn{2}{c}{BHE [MeV]} & & \multicolumn{2}{c}{Q$_{S}$ [eb]} && $g$-factor \\ 
\cline{5-6}\cline{8-9}
& & & & Th. & Expt. & & Th. & Expt. & & Th. \\
\hline
$^{164}$Gd & 6$^{+}$& $\pi$5/2$[532]\otimes$$\pi$7/2$[523]$   && 2.008   &          &&  4.178 &    && 1.259    \\
$^{164}$Gd & 6$^{-}$& $\pi$5/2$[532]\otimes$$\pi$7/2$[404]$   && 1.146   &          &&  4.278 &    && 1.007   \\
$^{164}$Gd & 6$^{-}$& $\nu$5/2$[512]\otimes$$\nu$7/2$[633]$   && 1.474   &          &&  4.127 &    && -0.126    \\
$^{164}$Gd & 6$^{+}$& $\nu$5/2$[642]\otimes$$\nu$7/2$[633]$   && 1.444   &          &&  4.214 &    && -0.144    \\
$^{164}$Gd & 12$^{-}$& \{$\pi$5/2$[532]\otimes$$\pi$7/2$[532]$\}
                $\oplus$\{$\nu$5/2$[512]\otimes$$\nu$7/2$[633]$\}  && 3.767   &      &&  5.060 &    && 0.584    \\
$^{164}$Gd & 12$^{-}$& \{$\pi$5/2$[532]\otimes$$\pi$7/2$[404]$\}
                $\oplus$\{$\nu$5/2$[642]\otimes$$\nu$7/2$[633]$\}  && 2.547   &      &&  5.295 &    && 0.439    \\
$^{164}$Gd & 12$^{+}$& \{$\pi$5/2$[532]\otimes$$\pi$7/2$[404]$\} 
                $\oplus$\{$\nu$5/2$[512]\otimes$$\nu$7/2$[633]$\} && 2.894   &      &&  5.186 &    && 0.449    \\

$^{166}$Gd & 6$^{+}$& $\pi$5/2$[532]\otimes$$\pi$7/2$[523]$   && 2.126   &          &&  4.228 &    && 1.267    \\
$^{166}$Gd & 6$^{-}$& $\pi$5/2$[532]\otimes$$\pi$7/2$[404]$   && 1.092   &          &&  4.328 &    && 1.013     \\
$^{166}$Gd & 6$^{-}$& $\nu$5/2$[512]\otimes$$\nu$7/2$[633]$   && 1.554   & 1.288\tnote{\emph{a}}   &&  4.251 &4.073\tnote{\emph{b}}  && -0.129    \\
$^{166}$Gd & 7$^{-}$& $\nu$7/2$[514]\otimes$$\nu$7/2$[633]$   && 1.816   &          &&  4.508 &    && 0.001     \\
$^{166}$Gd & 8$^{+}$& $\nu$7/2$[633]\otimes$$\nu$9/2$[624]$   && 2.234   &          &&  4.746 &    && -0.159     \\
$^{166}$Gd & 12$^{-}$& \{$\pi$5/2$[532]\otimes$$\pi$7/2$[523]$\}
                $\oplus$\{$\nu$5/2$[512]\otimes$$\nu$7/2$[633]$\} && 3.561   &         &&  5.217 &    && 0.585     \\
$^{166}$Gd & 12$^{+}$& \{$\pi$5/2$[532]\otimes$$\pi$7/2$[404]$\}
                $\oplus$\{$\nu$5/2$[512]\otimes$$\nu$7/2$[633]$\}  && 2.794   &         &&  5.342 &    && 0.448   \\
$^{166}$Gd & 13$^{+}$& \{$\pi$5/2$[532]\otimes$$\pi$7/2$[404]$\}
                $\oplus$\{$\nu$7/2$[514]\otimes$$\nu$7/2$[633]$\}  && 3.054   &         &&  5.417 &    && 0.481   \\

$^{168}$Gd & 6$^{+}$& $\pi$5/2$[532]\otimes$$\pi$7/2$[523]$   && 1.987   &          &&  4.167 &    && 1.266     \\
$^{168}$Gd & 6$^{-}$& $\pi$5/2$[413]\otimes$$\pi$7/2$[523]$   && 1.359   &          &&  4.268 &    && 1.017    \\
$^{168}$Gd & 7$^{-}$& $\nu$5/2$[512]\otimes$$\nu$9/2$[624]$   && 1.234  &           &&  4.524 &    && -0.135     \\
$^{168}$Gd & 6$^{+}$& $\nu$5/2$[512]\otimes$$\nu$7/2$[514]$   &&  0.866   &         &&  4.234 &    && 0.053     \\
$^{168}$Gd & 8$^{-}$& $\nu$7/2$[514]\otimes$$\nu$9/2$[624]$   &&  1.411   &         &&  4.726 &    && -0.017    \\
$^{168}$Gd & 12$^{-}$& \{$\pi$5/2$[413]\otimes$$\pi$7/2$[523]$\}
                $\oplus$\{$\nu$5/2$[512]\otimes$$\nu$7/2$[514]$\}  && 1.839   &      &&  5.320 &    && 0.545     \\
$^{168}$Gd & 12$^{+}$& \{$\pi$5/2$[532]\otimes$$\pi$7/2$[523]$\}
                $\oplus$\{$\nu$5/2$[512]\otimes$$\nu$7/2$[514]$\}  && 2.436   &      &&  5.195 &    && 0.679     \\
$^{168}$Gd & 13$^{-}$& \{$\pi$5/2$[532]\otimes$$\pi$7/2$[523]$\}
                $\oplus$\{$\nu$5/2$[512]\otimes$$\nu$9/2$[624]$\}  && 3.049   &       &&  5.308 &    && 0.527   \\
$^{168}$Gd & 13$^{+}$& \{$\pi$5/2$[413]\otimes$$\pi$7/2$[523]$\}
                $\oplus$\{$\nu$5/2$[512]\otimes$$\nu$9/2$[624]$\}  && 2.455   &       &&  5.435 &    && 0.402    \\

$^{166}$Dy & 6$^{+}$& $\pi$5/2$[532]\otimes$$\pi$7/2$[523]$   && 1.918   &          &&  4.286 &    && 1.254    \\
$^{166}$Dy & 6$^{-}$& $\pi$5/2$[532]\otimes$$\pi$7/2$[404]$   && 1.271   &          &&  4.386 &    && 1.007    \\
$^{166}$Dy & 6$^{-}$& $\nu$5/2$[512]\otimes$$\nu$7/2$[633]$   && 1.579   &          &&  4.226 &    &&-0.127    \\
$^{166}$Dy & 7$^{-}$& $\nu$7/2$[514]\otimes$$\nu$7/2$[633]$   && 1.845   &          &&  4.481 &    &&0.007    \\
$^{166}$Dy & 12$^{-}$& \{$\pi$5/2$[532]\otimes$$\pi$7/2$[523]$\}
                $\oplus$\{$\nu$5/2$[512]\otimes$$\nu$7/2$[633]$\}  && 3.775   &      &&  5.184 &    && 0.582    \\
$^{166}$Dy & 12$^{+}$& \{$\pi$5/2$[532]\otimes$$\pi$7/2$[404]$\}
                $\oplus$\{$\nu$5/2$[512]\otimes$$\nu$7/2$[633]$\}  && 3.107   &         &&  5.309 &    && 0.449  \\
$^{166}$Dy & 13$^{+}$& \{$\pi$5/2$[532]\otimes$$\pi$7/2$[404]$\}
                $\oplus$\{$\nu$7/2$[514]\otimes$$\nu$7/2$[633]$\}  && 3.371   &         &&  5.382 &    && 0.485  \\

$^{168}$Dy & 6$^{+}$& $\pi$5/2$[532]\otimes$$\pi$7/2$[523]$   && 2.039   &          &&  4.336 &    && 1.263   \\
$^{168}$Dy & 6$^{-}$& $\pi$5/2$[532]\otimes$$\pi$7/2$[404]$   && 1.218   &          &&  4.436 &    && 1.014   \\
$^{168}$Dy & 6$^{-}$& $\nu$5/2$[512]\otimes$$\nu$7/2$[633]$   && 1.314   &          &&  4.359 &    &&-0.130   \\
$^{168}$Dy & 7$^{-}$& $\nu$7/2$[514]\otimes$$\nu$7/2$[633]$   && 1.583   &          &&  4.623 &    && 0.006   \\
$^{168}$Dy & 8$^{+}$& $\nu$7/2$[633]\otimes$$\nu$9/2$[624]$   && 2.165   &          &&  4.866 &    && -0.160   \\
$^{168}$Dy & 12$^{-}$& \{$\pi$5/2$[532]\otimes$$\pi$7/2$[523]$\} 
                $\oplus$\{$\nu$5/2$[512]\otimes$$\nu$7/2$[633]$\}  && 3.233   &      &&  5.350 &    && 0.583  \\
$^{168}$Dy & 12$^{+}$& \{$\pi$5/2$[532]\otimes$$\pi$7/2$[404]$\}
                $\oplus$\{$\nu$5/2$[512]\otimes$$\nu$7/2$[633]$\}  && 2.675   &       &&  5.478 &    && 0.449 \\
$^{168}$Dy & 13$^{+}$& \{$\pi$5/2$[532]\otimes$$\pi$7/2$[404]$\}
                $\oplus$\{$\nu$7/2$[514]\otimes$$\nu$7/2$[633]$\}  && 2.942   &       &&  5.554 &    && 0.485 \\

$^{170}$Dy & 6$^{+}$& $\pi$5/2$[532]\otimes$$\pi$7/2$[523]$   && 1.902   &          &&  4.272 &    && 1.262     \\
$^{170}$Dy & 6$^{-}$& $\pi$5/2$[413]\otimes$$\pi$7/2$[523]$   && 1.483   &          &&  4.372 &    && 1.017     \\

$^{170}$Dy & 6$^{+}$& $\nu$5/2$[512]\otimes$$\nu$7/2$[514]$   && 0.881   & 1.644\tnote{\emph{c}}   &&  4.341 &    &&0.058    \\
$^{170}$Dy & 7$^{-}$& $\nu$7/2$[514]\otimes$$\nu$7/2$[633]$   && 1.352   &          &&  4.637 &    &&-0.136    \\
$^{170}$Dy & 8$^{-}$& $\nu$7/2$[514]\otimes$$\nu$9/2$[624]$   && 1.530   &          &&  4.845 &    &&-0.014    \\
$^{170}$Dy & 8$^{+}$& $\nu$7/2$[633]\otimes$$\nu$9/2$[624]$   && 2.008   &          &&  4.796 &    &&-0.157    \\
$^{170}$Dy & 12$^{-}$& \{$\pi$5/2$[413]\otimes$$\pi$7/2$[523]$\}
            $\oplus$\{$\nu$5/2$[512]\otimes$$\nu$7/2$[514]$\} && 1.965   &          && 5.452 &    && 0.549   \\
$^{170}$Dy & 12$^{+}$& \{$\pi$5/2$[532]\otimes$$\pi$7/2$[523]$\}
            $\oplus$\{$\nu$5/2$[512]\otimes$$\nu$7/2$[514]$\} && 2.354   &          && 5.327 &    && 0.680    \\
         
$^{170}$Dy & 13$^{-}$& \{$\pi$5/2$[512]\otimes$$\pi$7/2$[523]$\}
            $\oplus$\{$\nu$7/2$[514]\otimes$$\nu$7/2$[633]$\} && 3.076   &          &&  5.442 &    && 0.525    \\
$^{170}$Dy & 13$^{+}$& \{$\pi$5/2$[413]\otimes$$\pi$7/2$[523]$\}
            $\oplus$\{$\nu$7/2$[514]\otimes$$\nu$7/2$[633]$\} && 2.690   &          &&  5.569 &    && 0.403    \\           
\hline
\end{tabular}
}
\label{tab:isomer-prop}
\begin{tablenotes}
     \item[\emph{a}] Ref.~\cite{Patel2014}
     \item[\emph{b}] extracted from Ref.~\cite{Patel2014}
     \item[\emph{c}] Ref.~\cite{Soderstrom2016}
   \end{tablenotes}
  \end{threeparttable}
\end{table*}


The N=104 marks the midshell nuclei in this mass region, 
therefore rotational structure with large collectivity is expected \cite{Soderstrom}. Gd and Dy nuclei near N=104 are marked by
 large deformation for their ground band. Therefore, rotational bands are expected for these nuclei.
The yrast bands are shown in Fig.~\ref{fig:band-diagram}
for N=100, 102 and 104 isotopes of Gd (a-c) and Dy (d-f).  The  rotational structure and level spacing of
the yrast bands are well reproduced for the experimentally known cases. The yrast and excited band structures are obtained by mixing various K-configurations (based on particle-hole excitations across the proton and neutron Fermi surfaces)
with the HF ground state. In general, $K=0$ type 2p-2h and 4p-4h excitations are important for understanding the high-spin structures. 
This is one way of accounting paring by preserving particle number in HF formalism. For nuclei presently studied, however, these structures and also high-K structures mix very little, reflecting the robustness of
K-quantum number for axially symmetric nuclei.

K-isomers may arise from the breaking of one or more coupled nucleon pairs to form multiquasiparticle (multi-qp) states.
The single particle orbits near the proton and neutron Fermi surfaces are responsible for the configuration, excitation energy and 
other properties of the isomers. As shown in Fig.~\ref{fig:hforb}, 
the HF orbits near the neutron Fermi surface of  $^{168}$Gd are built on $h_{9/2}$ and $i_{13/2}$ orbits with admixtures from others. Excitation of a neutron from -5/2$[512]$ to 9/2$[624]$ gives the $K^{\pi}=7^{-}$ intrinsic structure.
Angular momentum projection (AMP) from this configuration gives the 2-qp band based on $K^{\pi}=7^{-}$ isomer. Similarly, the proton Fermi surface is built on $g_{7/2}$ and $h_{11/2}$ orbits with admixtures from others.
By exciting one proton from -5/2$[413]$ to 7/2$[523]$ one gets $K^{\pi}=6^{-}$ 2-qp isomer. The neutron and proton
2-qp configurations are coupled to give $K^{\pi}=13^{+}$ 4-qp structure. The high-K isomers and rotational bands based on the
isomers are shown in Fig.~\ref{fig:band-diagram}. Experimental observation of these high-K isomeric band are still awaited.The dominant configurations of the isomers and their energies along with
quadrupole moments as well as $g$-factor are listed in Table~\ref{tab:isomer-prop}.

Recently, K-isomers near $N=100$ nuclei have been observed for $^{162,164}$Sm~\cite{Patel2014,Yokoyama2017}, 
$^{163}$Eu~\cite{Yokoyama2017} and $^{164,166}$Gd~\cite{Patel2014,Yokoyama2017}. In Ref.~\cite{Patel2014},
Patel {\it et. al.} found an enhancement of collectivity near $Z\le$66, $N=100$ nuclei as predicted in Refs.~\cite{Ghorui2012,Satpathy2003}. Later, Yokoyama  {\it et. al.}~\cite{Yokoyama2017} also obtained similar results. The possible reason of this behaviour is the on-set deformation in these nuclei. The deformation ($\beta_{2}$) nearly saturates after $N=98$.
$K^{\pi}=4^{-}$ and $3^{+}$ 2-qp isomers are observed experimentally and explained well with the deformed Hartree-Fock 
model~\cite{Yokoyama2017}. In the present calculations we have considered, however, isomers with $K\ge6$.
In $^{164}$Gd, we predict a $K^{\pi}=6^-$ isomer at fairly low energy, around 1.5 MeV. This isomer is based on
neutron 2-qp excitation and the main configuration is assigned as $\nu5/2[512]\otimes\nu7/2[633]$.
Two 4-qp isomers with $K^{\pi}=12^-$ are also predicted. The 4-qp isomer involving two unpaired protons ($5/2^{-},7/2^{+}$) and
two unpaired neutrons ($5/2^{+},7/2^{+}$) is energetically favoured (with
excitation energy about 2.5 MeV) compared to the other $K^{\pi}=12^-$ state which lies about 1.2 MeV higher in energy. We also predict a $K^{\pi}=12^+$ state at relatively higher excitation energy above 3.0 MeV.

A $K^{\pi}=6^{-}$ isomeric state at 1.288 MeV is experimentally observed for $^{166}$Gd~\cite{Patel2014}. This isomer has the 
configuration  $5/2[413]\otimes7/2[523]$. In DHF calculations, we obtained the isomeric state at 1.554 MeV, around 260 keV higher in energy. 
Our DHF results agree reasonably with the experimental value. Moreover, from the 
quoted $\beta_{2}$ value in Ref.~\cite{Patel2014}, we extracted the spectroscopic 
quadrupole moment ($Q_s$) using the formula  of Ref.~\cite{Raman2001}.
The experimental extracted value, which is 4.073 $eb$ is nicely reproduced by DHF calculations (4.251 $eb$).
A $K^{\pi}=6^{-}$ configuration due to proton excitation
from $-5/2^{-}$ to $7/2^{+}$ state is obtained at 1.092 MeV. The
$K^{\pi}=6^{+}, 7^{-}$ and $8^{+}$ states based on proton excitations
with 2-qp nature are predicted relatively higher excitation energies 
about 2.0 MeV. 4-qp configurations based on two proton and two neutron excitations are calculated to lie $\approx$3.0 MeV. All these high-K configurations are listed in Table.~\ref{tab:isomer-prop}.

In $^{168}$Gd, a 2-quasineutron isomer with configuration 
$\nu5/2[512]\otimes\nu7/2[514]$
is obtained at very low-excitation energy $E\approx$870 keV. The
low-excitation energy of this isomer can be understood as the neutron number increases from $^{164}$Gd to $^{168}$Gd, the $\nu7/2[514]$ orbital pushed down near to the neutron Fermi surface and can be easily 
occupied by excitations. This $K^{\pi}=6^{+}$ isomer possibly have longer 
half-life because of large hindrance factor. Also a $K^{\pi}=7^{-}$ isomer from neutron excitation is predicted at 1.234 MeV. The main 
configuration of this isomer is $\nu5/2[512]\otimes\nu9/2[624]$. 
We also predict 4-qp isomers with $K^{\pi}=12^{-}$(1.839 MeV), $K^{\pi}=12^{+}$(2.436 MeV) and $K^{\pi}=13^{+}$(2.455 MeV) at fairly low energy for this nucleus (as shown in Table~\ref{tab:isomer-prop}).
 These 4-qp isomers are based on two unpaired neutron and two unpaired proton configurations those have low excitation energies. 
Therefore, 4-qp isomers also obtained with low excitation energies.
As in case of Gd nuclei, 2-quasineutron $K^{\pi}=6^{-}$ ($\nu5/2[512]\otimes\nu7/2[633]$ ) isomer
is predicted for $^{166,168}$Dy nuclei at 1.579 MeV and 1.314 MeV, respectively. From proton excitation, $K^{\pi}=6^{+}$
isomer at E$\sim$ 2.0 MeV with configuration $\pi5/2[532]\otimes\pi7/2[523]$ is also obtained for both N=100 and N=102 isotopes of Dy. Experimental observations of these K-isomers are
awaited.

The midshell isotope $^{170}$Dy (Z=66 and N=104) is a very interesting candidate for low-lying isomers (both 2-qp and 4-qp).
However, spectroscopic properties for the yrast band as well as excited states are extremely limited from experimental point of view.
Regan {\it et. al.,}~\cite{Regan2002} from potential energy surface calculations predicted a $K=6^{+}$
  2-quasineutron isomeric state (based on the 5/2[512] and 7/2[514] orbitals) at E$\approx$1.2 MeV in $^{170}$Dy.
 Recently, this isomer is measured experimentally and determined to have an excitation energy of 1643.91(23) keV~\cite{Soderstrom2016}.
 Our DHF calculations give the excitation energy of the $K=6^{+}$ isomer 
 about $\sim$1.0 MeV. 
 
 The neutron orbitals 5/2[512] and 7/2[514] which are involved in the $K=6^{+}$
 isomer are not strongly sensitive to the deformation and lie close
 to the Fermi surface. 
 These orbitals evolve in a similar way with deformation. 
 It is, therefore, very unlikely that the uncertainties 
 in the deformation can be a dominant factor for the energy difference. Usually mixing of $K=0$ configurations of $2p-2h$ and $4p-4h$ paring type
 excitaions are found to be important for the excitation energies of high-K structure. But, as shown in Table~\ref{tab:hf-prop}, the interaction energies ($V_{pp}, V_{nn}~\text{and}~V_{pn}$) are very 
 large and energy gain from further mixing of other K-configurations is negligible (100 keV in this case). Also, as shown in the relative binding energy plot in Fig.~\ref{fig:be-rel}, 
 most of the important correlations within the present model space are already taken care of, therefore, the extension of the present configuration space may be helpful to understand the energy difference. This is beyond the scope of the present study and we plan to investigate in future.

 A $K=6^{+}$ configuration involving two unpaired protons with intrinsic
 structure $\pi5/2[532]\otimes\pi7/2[523]$ is also obtained. But this lies
 higher in excitation energy $\sim$1.9 MeV. This configuration does not mix 
 much ($<$2\%) with the 2-quasineutron $K=6^{+}$ configuration.
 Another 2-qp  isomer ($K^{\pi}=7^{-}$) is identified at 1.352 MeV with configuration $\nu7/2[514]\otimes\nu7/2[633]$.
 We also observed 4qp isomers with $2\pi\otimes 2\nu$ structures with excitation energy below 3.0 MeV. These isomers and their properties are given in Table~\ref{tab:isomer-prop}. Some of them are predicted at low excitation energies.

\begin{figure}[ht!]
\includegraphics[width=0.48\textwidth]{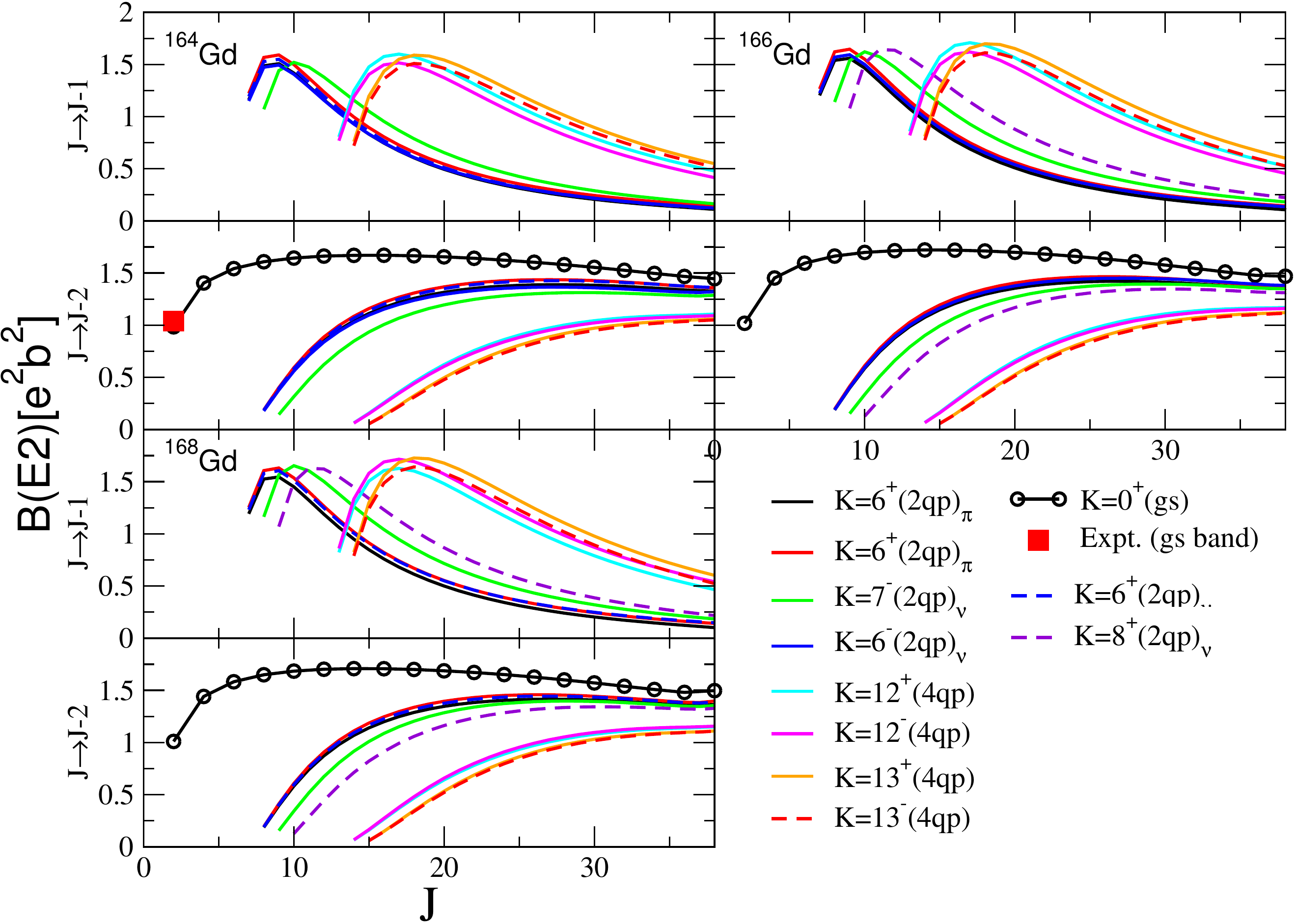}
\caption{\label{fig:be2gd}
B(E2) values for ground as well as 2qp and 4qp bands for $^{164,166,168}$Gd isotopes. Experimental data taken from Ref.~\cite{Singh2018}.}
\end{figure}

For completeness, we also study the transition probabilities of the ground band as well as bands based on the isomers. Electromagnetic transition probabilities are important quantities to test the collectivity of the nuclear states. The comparison of calculated electromagnetic properties (e.g, B(E2), B(M1), g-factor etc.) with experimentally available data ensures the reliability of the wave functions used to obtain them. Unfortunately, due to experimental difficulties, information on electromagnetic
properties are very scarce for the nuclei presently studied. 
 
The B(E2) values for a transition from an initial state $\alpha J_{1}$ to final state $\beta J_{2}$  is given by
\begin{equation}
B(E2; \alpha J_{1} \to \beta J_{2})=\frac{1}{(2 J_{1} + 1)}\left | \sum_{i=p,n} \langle \Psi_{K_2}^{\beta J_2}|| Q_{2}^{i}|| \Psi_{K_1}^{\alpha J_1} \rangle \right |^{2}
\end{equation}
where $i=p$ and $n$ for protons and neutrons, respectively. The summation is for quadrupole moment operators of protons and
neutrons.
\begin{figure}[h!]
\includegraphics[width=0.48\textwidth]{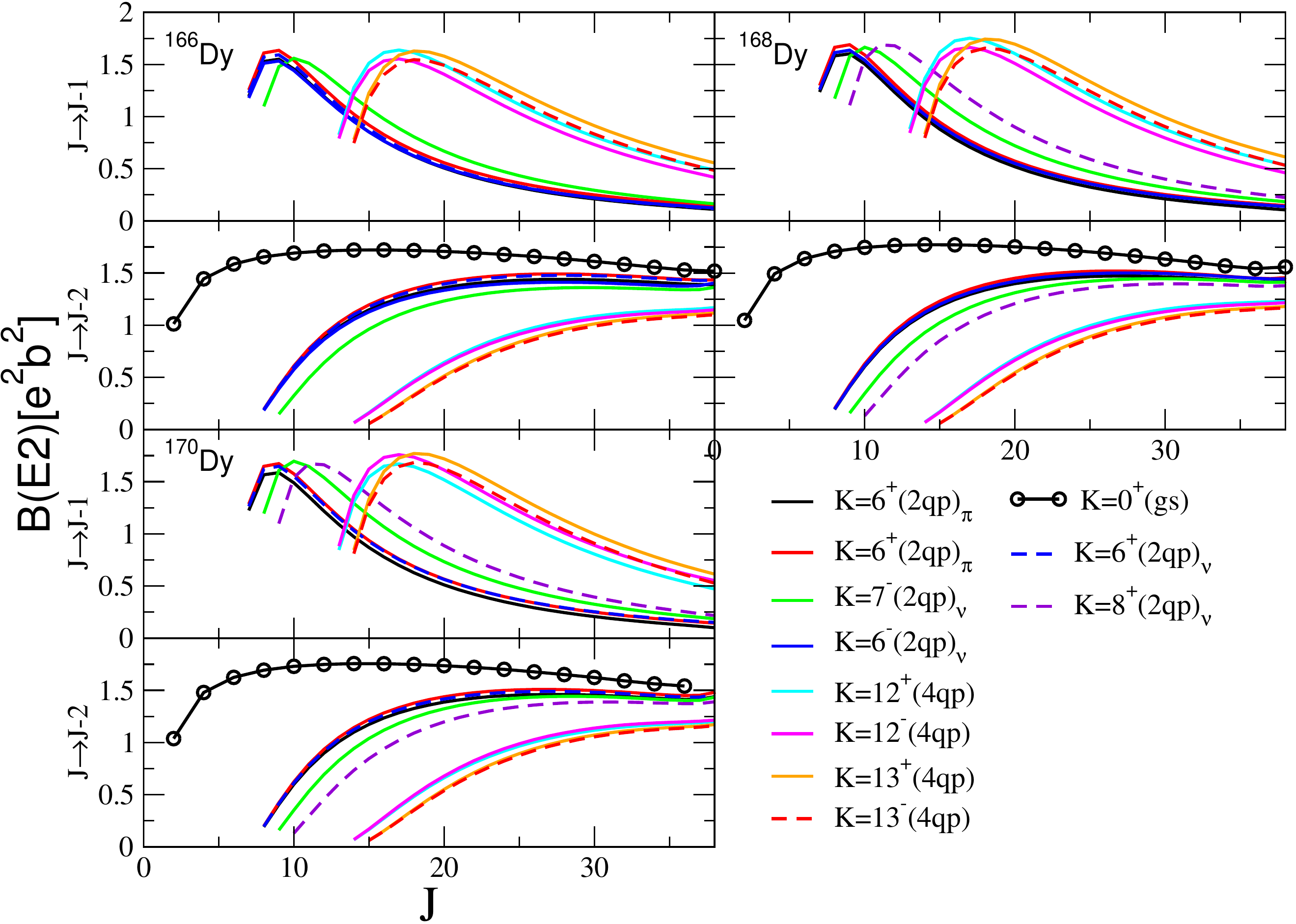}
\caption{\label{fig:be2dy}
Same as Fig.~\ref{fig:be2gd}, but for $^{166,168,170}$Dy isotopes.}
\end{figure}
The B(E2) transition probabilities are depicted in Fig.~\ref{fig:be2gd} and Fig.~\ref{fig:be2dy} for Gd and Dy isotopes, respectively. The proton effective charges $e_{\pi}=(1 +\frac{Z}{A})e$ and
for the neutron $e_{\nu}=(2.1\times\frac{Z}{A})e$ are used \cite{Raman2001}. The $B(E2;2^{+}_{1} \to 0^{+})$
for $^{164}$Gd is only known from experiment and measured to be 198$^{+11}_{-9}$ W.u. \cite{Singh2018}. We calculate it to be 170 W.u., which compares well with the experimental value. 
In Figs.~\ref{fig:be2gd} and \ref{fig:be2dy}, the $B(E2;J \to J-2)$ values for the ground
band show similar trend throughout the isotopic chains studied 
here.
The smooth variation of transition probabilities support the rotational structure in these deformed nuclei for the ground bands.

For K-isomeric bands, in case of $J \to J-2$ transitions, at low-spins $B(E2)$ values are small. The values increase with $J$
and at high-spin become almost constant. For $J \to J-1$ 
transitions, the $B(E2)$ values increase rapidly with $J$ and 
attain a maximum value of $\approx$1.5 $e^{2}b^{2}$ then gradually 
decrease. The maximum value occurs at $J\sim$10$\hbar$ for 2-qp bands and $J\sim$18$\hbar$ for 4-qp band for both Gd and Dy isotopic chains. 

\begin{figure*}[h!]
\includegraphics[width=1.0\textwidth]{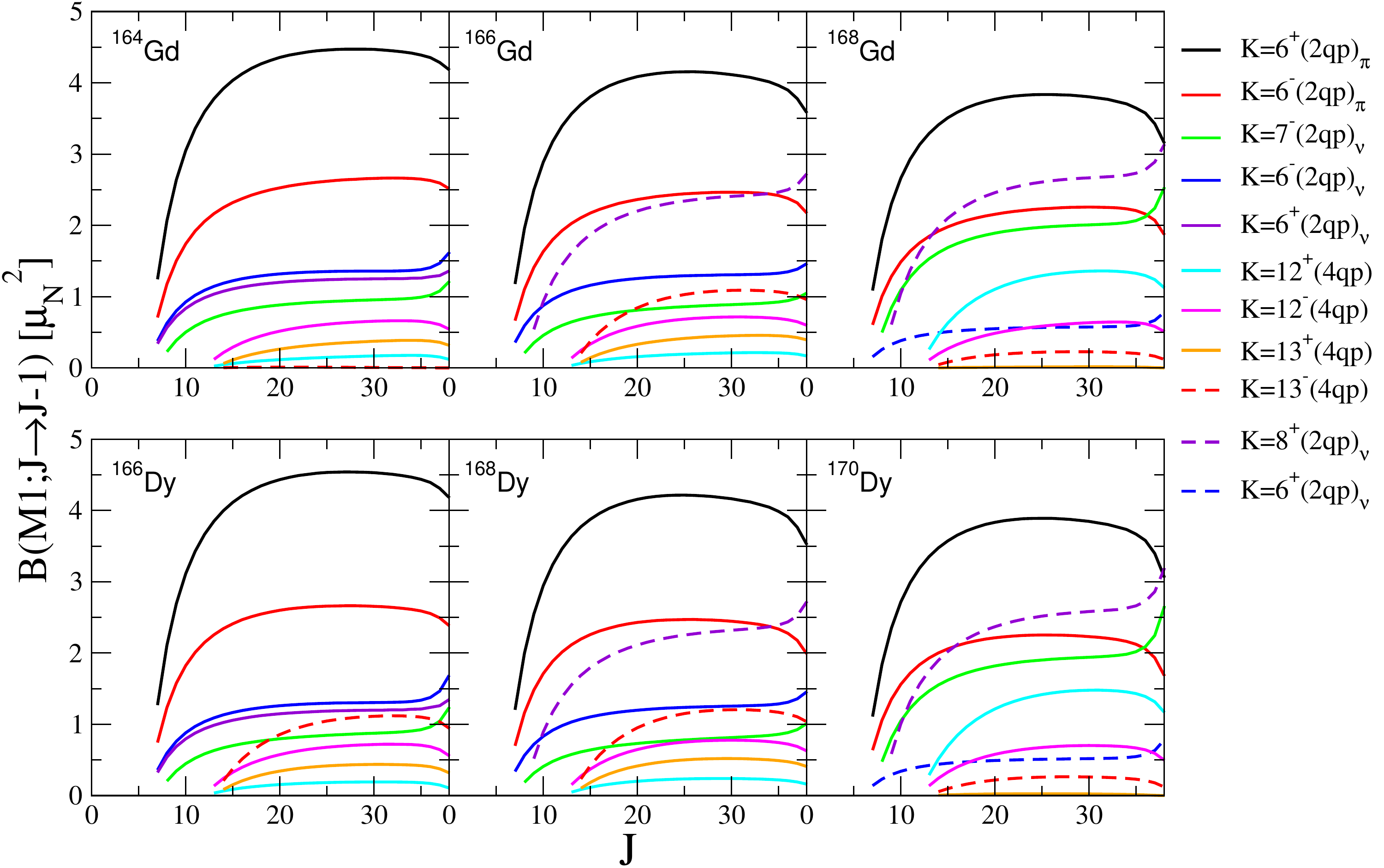}
\caption{\label{fig:bm1} BM1 transitions for Gd (upper panel) and Dy (lower panel) nuclei.}
\end{figure*}

The reduced magnetic dipole (M1) transition moments between
initial and final states are given by

\begin{strip}
\begin{eqnarray}
B(M1; \alpha J_{1} \to \beta J_{2})=\frac{3}{4\pi(2 J_{1} + 1)} \left | \sum_{i=p,n} 
\langle \Psi_{K_2}^{\beta J_2}|| g_{l}^{i}l_{i} + g_{s}^{i}s_{i} || \Psi_{K_1}^{\alpha J_1} \rangle \right |^{2}
\end{eqnarray}
\end{strip}
\noindent where $i=p$ and $n$ for protons and neutrons, respectively. The quantities $g_{l}$ and $g_{s}$ represents orbital and spin g-factors, respectively.

In Fig.~\ref{fig:bm1}, we have shown the B(M1) values for isomeric bands. The g-factors of $g_{l}=1.0 \mu_{N}$ and $g_{s}=5.586\times0.75 \mu_{N} $ for protons and 
$g_{l}=0 \mu_{N}$ and $g_{s}=-3.826\times0.75 \mu_{N}$ for neutrons are used for the
calculations. The quenching of spin g-factors by 0.75 is taken in account to consider
the core polarization effect~\cite{Castel1990}. As these nuclei have similar deformation,
therefore the bands with similar quasi-particle configuration show nearly same trends. The large M1 transition probabilities for
$K=6^{+}$ band with configuration $\pi$5/2[532]$\otimes\pi$7/2[523] are noticeable for all nuclei studied here. Mainly the proton part contributes for this band. For most of the 4-qp bands,
contribution from orbital part is negligible; 
proton and neutron spin parts add-up destructively 
(because of negative value of the neutron spin g-factor) to give a small net quantity except for 
$K^{\pi}=12^{+}$  bands in N=104 isotone.

\section{Summary}
High-K isomers for the midshell Gd and Dy isotopes are studied with the 
self-consistent Deformed Hartree-Fock model. Angular momentum projection 
is performed to study bands built on the ground and excited intrinsic 
configurations and K-isomers. The deformation properties of these nuclei 
are discussed. The $\beta_{2}$ deformation varies smoothly beyond N=96 with maximum around N=102 for these nuclei. The known isomers are well reproduced in the present calculations. 
Electromagnetic properties (quadrupole moment and $g$-factor) of the isomers are also calculated. From the systematics, 
2-qp and 4-qp isomers at low-excitation energies are
predicted for these isotopes and their main-configuration as well as electromagnetic 
properties, which are important for experiments, are listed. These high-K isomers give unique access to the high-spin spectroscopy  of {\it hard-to-reach} nuclei in the neutron-rich deformed rare-earth nuclei. Experimental information about high-spins for these neutron rich nuclei are very sparse and our results for the high-spins give some insight and trends. The present systematical study will further aid for the experimental quest on high-spin structure with the current generation of facilities.

\section{Acknowledgments}
Research at SJTU was supported by the National Natural
Science Foundation of China (No. 11135005), by the 973 Program
of China (No. 2013CB834401), and by the Open Project
Program of State Key Laboratory of Theoretical Physics, Institute
of Theoretical Physics, Chinese Academy of Sciences,
China (No. Y5KF141CJ1). CRP acknowledges the support of the 
Department of Science and Technology, 
Govt. of India (DST Project SB/S2/HEP-006/2013) during this work.

\bibliographystyle{epj}
\bibliography{rare-earth}


\appendix
\renewcommand{\thesection}{\Alph{section}~}
\renewcommand{\thesubsection}{\arabic{subsection}}
\setcounter{table}{0}
\setcounter{figure}{0}
\setcounter{footnote}{0}
\setcounter{equation}{0}
\renewcommand{\theequation}{\thesection\arabic{equation}}
\renewcommand{\thefigure}{\thesection\arabic{figure}}
\makeatletter
\renewcommand\@biblabel[1]{[\thesection#1]}
\makeatother
\section{Deformed Hartree-Fock equation with axial symmetry}\label{appn}
The configuration generated in deformed Hartree-Fock theory are obtained 
after self-consistent calculation including the effect of the residual interaction.
The Hamiltonian is 
\begin{equation}
    H=H_{0} + V~,
\end{equation}
where, $H_{0}$ obtained from the single particle energies of protons and neutrons
$H_{0}=\sum_{jm}\varepsilon_{j}a^{\dagger}_{jm}a_{jm}$. $V$ is the residual interactions
among valance nucleons $V_{pp}$, $V_{pn}$ and $V_{nn}$. The matrix elements of $V$'s are
suitably antisymmetrized for $V_{pp}$ and $V_{nn}$.

With axial-symmetry of the Hartree-Fock (HF) field a HF nucleon orbit $\mid\eta m\rangle$ is
a superposition of spherical orbits:
\begin{equation}
\mid \eta m\rangle =\sum_{j}C^{\eta}_{jm}\mid jm\rangle \,
\end{equation}
where j is the angular momentum of the shell model single particle states and  
$m$ its projection. The amplitude $C^{\eta}_{jm}$ of the spherical state 
$\mid jm\rangle$ in the orbit $\eta$ obtained by solving the deformed
Hartree-Fock equations \thinspace
\begin{equation}
E^{\eta }C_{jm}^{\eta }=\varepsilon _{j}C_{jm}^{\eta} + 
\sum V(j_{1}mj_{2}m_{2};jmj_{4}m_{2})\rho
_{j_{2}m_{2}j_{4}m_{2}}\,C_{j_{1}m}^{\eta}\,
\label{eq:hf}
\end{equation}
where $\varepsilon_{j}$ the single-particle energy of shell model  orbit
$j(\equiv nlj)$, $V$ denotes the two-body interaction among nucleons and $\rho$ is the
density matrix\thinspace
\begin{equation}
\rho _{j_{1}mj_{2}m}=\sum_{occupied}C_{j_{1}m}^{\eta~*}C_{j_{2}m}^{\eta} \,
\label{eq:rho}
\end{equation}
Eqns.~\ref{eq:hf} and \ref{eq:rho} are solved by iteration process till there is convergence, by diagonalizing the HF Hamiltonian $h=\varepsilon + V\rho=\varepsilon+\Gamma$ in each iteration. $\Gamma=V\rho$ is the HF self-energy.

All intrinsic quantities follow from the density
matrix $\rho$ of the HF calculation. For example,
\begin{equation}
    \langle Q_{20} \rangle_{HF}=Tr({Q_{20}\rho})
\end{equation}
So deformation is not something externally imposed.
It follows dynamically from the HF theory. 
Of course, the intrinsic states are not the physical states. One needs angular momentum 
projection from the intrinsic states to obtain 
the physical states with good angular momentum.

The effects of residual interaction $V$ is taken into account
at each iteration of the HF procedure (Eqns.~\ref{eq:hf} and \ref{eq:rho}) and it plays important role in determining the mixing amplitudes in the HF process. Essentially each HF configuration shows the intrinsic distribution of the valence nucleons in the deformed shell model orbits. It is found that angular-momentum projection from a few low-lying HF configurations gives a reasonable description of the yrast structure. This is due to the fact that the residual interaction has been used in the solution of the HF equations, so that the HF single-particle states and the various multiparticle configurations built from them are already closer to the final answer that comes from a full solution of the many-body Schr\"{o}dinger equation~\cite{Macfarlane1971,Khadkikar1971}. In fact angular momentum projection from a single K-configuration reproduces the energy features of the low-lying yrast states with good accuracy. Therefore, further elaborate configuration mixing (band-mixing)
is not necessary. 

\end{document}